\documentclass[a4paper]{jpconf}
\usepackage{graphicx}
\usepackage{epsf}

\usepackage[]{latexsym}
\usepackage{graphicx}
\usepackage{graphics}
\usepackage{amsmath,amssymb} %,bbm}
\usepackage{subeqnarray}
\usepackage{slashed}							
\newcommand{\be}{\begin{equation}}\newcommand{\ee}{\end{equation}}
\newcommand{\bea}{\begin{eqnarray}}\newcommand{\eea}{\end{eqnarray}}
\newcommand{\brr}{\begin{array}}\newcommand{\err}{\end{array}}
\newcommand{\bit}{\begin{itemize}}\newcommand{\eit}{\end{itemize}}
\newcommand{\ben}{\begin{enumerate}}\newcommand{\een}{\end{enumerate}}

\newcommand{\ba}{\begin{array}}
\newcommand{\ea}{\end{array}}

\newcommand{\arccotg}{\mathop{\operator@font arccotg}\nolimits}

\def\rar{\rightarrow}

\def\bt{\beta}

\def\1{{_{1}}}\def\2{{_{2}}}

\def\noHe0{:\;\!\!\;\!\!:H_e(0):\;\!\!\;\!\!:}
\def\noHm0{:\;\!\!\;\!\!:H_\mu(0):\;\!\!\;\!\!:}

\def\rar{\rightarrow}

\def\bt{\beta}

\def\1{{_{1}}}\def\2{{_{2}}}

\begin{document}
\begin{flushright}
KCL-PH-TH/2015-05
\end{flushright}
\title{Noncommutative spectral geometry, Bogoliubov transformations and neutrino oscillations  }

\author{Maria Vittoria Gargiulo}
\address{Dipartimento di Fisica "E.R.Caianiello", I.N.F.N., Universit\`a di Salerno, I-84100 Salerno, Italy}
\ead{mgargiulo@unisa.it}

\author{Mairi Sakellariadou }
\address{Department of Physics, King's College, University of London, Strand WC2R 2LS, London, U.K.}
\ead{mairi.sakellariadou@kcl.ac.uk}

\author{Giuseppe Vitiello}
\address{Dipartimento di Fisica ``E.R.Caianiello'', I.N.F.N., Universit\`a di Salerno,
I-84100 Salerno, Italy}
\ead{vitiello@sa.infn.it}

\begin{abstract}
  In this report we show that neutrino mixing is intrinsically
  contained in Connes' noncommutative spectral geometry construction,
  thanks to the introduction of the doubling of algebra, which is
  connected to the Bogoliubov transformation. It is known indeed that
  these transformations are responsible for the mixing, turning the
  mass vacuum state into the flavor vacuum state, in such a way that
  mass and flavor vacuum states are not unitary equivalent. There is
  thus a red thread that binds the doubling of algebra of Connes'
  model to the neutrino mixing.
\end{abstract}

\section{Introduction}

In the noncommutative spectral geometry (NCSG)
construction~\cite{ncg-book1}-\cite{Capozziello}, introduced by Alain
Connes, non commutative geometry is combined with a spectral triples
so to obtain the full Lagrangian of the Standard Model (SM), minimally
coupled with gravity and compatible with neutrino mixing.  The
coupling with gravity is obtained thanks to the fact that this
construction uses a group of symmetry which encodes both, the
diffeomorphism which control general relativity, and the local gauge
invariance which rules the gauge theories the SM is based on.

The key point in Connes' construction is the doubling of the space and of the algebra.
It has been shown~\cite{Sakellariadou:2011wv} that this doubling is related to the gauge field structure,  to dissipation and that the NCSG classical construction carries in itself the seeds of quantization. Closely following ref.~\cite{GSV}, in this report we show that the doubling of the algebra is also the main mathematical tool on which neutrino mixing is based.

A brief summary of the NCSG construction is presented in Section 2. The doubling of the algebra and the Bogoliubov transformations are discussed in Sections 3 and 4 and the neutrino mixing is studied in Section 5. Section 6 is devoted to the conclusions.

\section{A brief summary of NCSG construction}
The skeleton of the NCSG construction for the Standard Model is presented below.

Considering the action functional $S$  at low energy scale,  $S = S_{\cal E-H} + S_{\cal SM} $, which is the sum of the Einstein-Hilbert action ($ S_{\cal E-H} $) and SM action ($ S_{\cal SM} $), it can be noticed that the two parts do not share the same symmetries. The former is ruled by outer automorphism invariance (diffeomorphism) the latter by inner automorphism (local gauge transformation).Connes considers then a model of a two-sheeted space, realized as the product of a four dimensional smooth compact Reimannian manifold $ {\cal M} $ with a fixed spin structure by a discrete noncommutative space $ {\cal F} $ composed by only two points. The geometric space is thus defined as the tensor product of the continuous geometry of ${\cal M}$ by an internal geometry ${\cal F}$.

In this approach, the SM is seen as a phenomenological model with the geometry of space-time such that the Maxwell-Dirac action functional leads to the SM action.
The noncommutative nature of the discrete space ${\cal F}$ is expressed by the real
spectral triple ${\cal F}=({\cal A}_{\cal F}, {\cal H}_{\cal F},
D_{\cal F})$, where ${\cal A}_{\cal F}$  is an involution of operators on the finite-dimensional Hilbert space ${\cal H_F}$ of Euclidean fermions,
and ${\cal D_F}$ is a self-adjoint unbounded operator in
${\cal H}_{\cal F}$.  The spectral nature of the
triple implies that the Dirac operator $D_{\cal
F}$ of the internal space is the fermionic mass matrix. The
Dirac operator is the inverse of the Euclidean propagator
of fermions. The
information  carried by the metric are contained in the algebra ${\cal A}_{\cal F}$.
The spectral nature approach is intrinsic to the the noncommutative spectral geometry.
The  ${\cal M} \times {\cal F}$ product geometry is specified by the rules:
\bea
&{\cal A}=C^\infty ({\cal M}) \otimes {\cal A_F} = C^\infty ({\cal M , A_F}),\\
&{\cal H} = {\cal L}^2({\cal  M }, S) \otimes {\cal H_F} =  {\cal L}^2({\cal M} , S \otimes {\cal H_F) }\\
& D = \slashed {\partial}_{{\cal M}} \otimes 1 + \gamma_5 \otimes D_{ {\cal F}},
\eea
where ${\cal L}^2({\cal  M }, S)$ is the space of square integrable Dirac spinors over ${\cal  M }$, the Dirac operator on ${\cal  M }$ is denoted by $\slashed {\partial}_{{\cal M}}= i \gamma^\mu \nabla^s_\mu$ (with $\nabla^s_\mu$ is the spin connection $\nabla^s_\mu =\partial_\mu + \frac 1 2 \omega^{ab}_\mu \gamma_{ab}$), $\gamma_5$ is the chirality
operator in the four-dimensional case,
$C^\infty ({\cal M}) $ is the algebra of smooth functions on  ${\cal M}$ acting on ${\cal H_F}$  as simple multiplication operators.
Within the noncommutative spectral geometry $D$ plays the role of the inverse of the line element $ds$.

The main results of the model are obtained by using the spectral action principle, which is of the form
\begin{equation}
Tr(f(\frac {\cal D} \Lambda)) ,
\end{equation}
where $\cal D$ denotes the
inverse of the Dirac operator, $\Lambda$ fixes the energy scale and f is a cut-off positive even function of the real variable; it
falls to zero for large values of its argument. The function f only plays a role through its momenta $f_0$, $f_2$ and $f_4$, where $f_k ~=~ \int_0^{\infty} f(u) u^{k-1}\ du$ for $ k > 0 $ and $f_0 = f(0)$ , which are three parameters of the model and are related to the coupling constant at unification, the gravitational constant and the cosmological constant, respectively.
Its asymptotic expansion is:
\begin{equation}
Tr(f(\frac {\cal D} \Lambda)) \sim 2 \Lambda^4 f_4 a_0 + 2 \Lambda^2 f_2 a_2 + f_0 a_4 .
\end{equation}
The action functional, applied to inner fluctuations, only accounts for the bosonic part of the model.
The term in $\Lambda^4$ gives the cosmological term, the term in $\Lambda^2$ gives the Einstein-Hilbert action functional, and the $\Lambda$-indipendent term yields the Yang-Mills action for the gauge fields.
The coupling with fermions is obtained by adding the term $ \frac 1 2 <J \psi , {\cal D} \psi>$, where J is a real structure on the spectral triple and $\psi$ is a spinor in the Hilbert space ${\cal H_F}$ of the quarks and leptons. The computation of the spectral action functional gives the full Lagrangian for the Standard Model minimally coupled with gravity, with neutrino mixing and Majorana mass terms. The vacuum expectation value of the Higgs field is related to the non commutative distance between the two sheets.  In the model neutrino mixing is obtained in analogy with the quarks case.

\section {Doubling the Algebra and the deformed Hopf algebra coproduct}
In this Section we show that Bogoliubov transformations are ``built in'' in the algebra doubling in the NCSG. This implies in turn that the the NCSG construction insists on a collection of unitarily inequivalent Hilbert spaces which are related among themselves (phase transitions) by Bogoliubov transformations. Since these also characterize the mixing transformations of neutrinos, the NCSG construction also contains implicitly the neutrino mixing phenomenon.

Consider the Dirac spectral triple $({\cal A}_1, {\cal H}_1,
{\cal D}_1)=(C^{\infty}({\cal M},{\cal L}^2({\cal M},S), \slashed {\partial}_{{\cal M}})$
and its product with the finite geometry $({\cal A}_2, {\cal H}_2,
{\cal D}_2) = ({\cal A}_{\cal F}, {\cal H}_{\cal F}, {\cal D}_{\cal F})$. The product geometry ${\cal M}\times {\cal F}$ is given by
\bea\label{double}
{\cal A} &=& {\cal A}_1 \otimes {\cal A}_2 ~,~{\cal H} = {\cal H}_1 \otimes {\cal H}_2~,\\ \nonumber
{\cal D} &=& {\cal D}_1 \otimes 1 + {\gamma}_1 \otimes {\cal D}_2~,\\ \nonumber
\gamma &=& {\gamma}_1 \otimes {\gamma}_2 J=  J_1 \otimes J_2~,\nonumber
\eea
with
\be \label{J} J^{2} = -1, ~~ [J,{\cal D}] = 0 , ~~[J_1,\gamma_1] = 0 ,~~\{J,\gamma\}= 0,
\ee
where $J$ is an antilinear isometry and commutators and anticommutators are denoted by square and curl brackets, respectively.
In the formalism of the algebra doubling an important r\^ole is played by the noncommutative \textit \itshape{q}-deformed Hopf algebra, pointing to a deep physical meaning of the noncommutativity in this construction.
Indeed, the map  $ {\cal A} \to {\cal A}_1 \otimes {\cal A}_2 $ is just the Hopf coproduct map  $ {\cal A} \to {\cal A} \otimes {\bf 1} + {\bf 1} \otimes {\cal A} $ .
Actually for a noncommutative algebra the coproduct is defined by
\bea
\label{eqn:copr}
&\Delta a_q = a_q  \otimes q^H + q^{-H} \otimes a_q \\
&\Delta a^{ \dagger}_q = a^{ \dagger}_q \otimes  q^H + q^{-H}  \otimes a^{ \dagger}_q  .
\eea
The fermionic algebra ${h}(1|1)$ is generated by a set of operators $\{a , a^{ \dagger},H, N \} $ with anticommutation relations:
\bea
&\{ a , a^{\dagger} \} = 2H , \quad
 [ N , a ] = -a, \quad
[ N , a^{\dagger} ] = a^{\dagger},\quad
[ H , \bullet ] = 0 .
\eea
The deformed algebra $h_q(1\mid 1)$ relations are
\bea
&\{ a_q , a^{\dagger}_q \}= [2H]_q ,\quad
[ N , a_q ] = -a_q , \quad
[ N , a^{\dagger}_q ] = a^{\dagger}_q ,\quad
[ H , \bullet ] = 0,
\eea
where $N_q\equiv N$ and $H_q\equiv H$. The Casimir operator ${\cal C}_q$ is given by ${\cal C}_q = N[2H]_q - a^{\dagger}_q a_q$ , where
\[
[x]_q=\frac {q^x -q^{-x}} {q-q^{-1}} .
\]
and together with the coproduct defined in  Eqs.(\ref{eqn:copr}) we have
\bea
&\Delta H = H  \otimes {\bf 1} + {\bf 1} \otimes H ,\\
&\Delta N = N  \otimes  {\bf 1} + {\bf 1} \otimes N  .
\eea
Requiring that $a$ and $a^{\dagger}$, and $a_q$ and $a^{\dagger}_q$ are adjoint operator implies that $q$ can only be real or of modulus one; for fermion, in fact, $q\sim e^{i \theta} $ .
Considering a two-mode Fock space $F_2 = F_1 \otimes F_1 $ it is possible to write
\bea
\Delta a_q  &=&  a_1 q^{\frac 1 2 }   +  q^{-\frac {1} 2 } a_2, \\
\Delta a^{ \dagger}_q  &=&  a^{ \dagger}_1 q^{\frac 1 2 }  + q^{-\frac {1} 2 }  a^{ \dagger}_2,  \\
\Delta N &=& N_1 + N_2, \quad \Delta H =1.
\eea

One key point is that the noncommutative Hopf coproduct turns out to be strictly related to the Bogoliubov transformations~\cite{Celeghini:1998sy}.
It is indeed possible to identify $ a_1 \equiv a$ , $a^{\dagger}_1 \equiv a^{\dagger}$ , and  $ a_2 \equiv \tilde{a}$ , $a^{\dagger}_2 \equiv \tilde{a}^{\dagger}$, where $\tilde{O}$, with $O$ a generic operator, denotes the "tilde-coniugation" operation defined by the ``tilde-coniugation'' rules~\cite{Umezawa1983}:
\bea
&\widetilde{(OO')}=\tilde{O}\tilde{O'},\\
&\widetilde{(\alpha O+\beta O')}=\alpha^\ast \tilde{O}+\beta^\ast\tilde{O'},\\
&\widetilde{(O^{\dagger})}=\tilde{O}^{\dagger},\\
&\widetilde{(\tilde{O})}= O,\\
&\{O,\tilde{O'}\}=\{O,\tilde{O'}^{\dagger}\}=0 ,\\
&\widetilde{|O\rangle}=|O \rangle.
\eea
It is now possible to define the new operators $A_q$ and $B_q$ as follow
\bea
& A_q\equiv \frac{\Delta a_q}{\sqrt{[2]_q}}=\frac 1 {\sqrt{[2]_q}} (e^{i\theta}a+e^{-i\theta}\tilde{a}),  \\
& B_q\equiv \frac 1 {i \sqrt{[2]_q}} \frac \delta {\delta\theta} \Delta a_q=\frac {2q} {\sqrt{[2]_q}} \frac \delta {\delta q} \Delta a_q =\frac 1 {\sqrt{[2]_q}} (e^{i\theta}a-e^{-i\theta}\tilde{a}).
\eea
and h.c., with $q=q(\theta)\equiv e^{i2\theta}$.
The anti-commutation relation are:
\bea
&\{A_q,A^{\dagger}_q\}=1, \quad
\{B_q,B^{\dagger}_q\}=1 , \quad
\{A_q,B_q\}=0, \\
&\{A_q,B^{\dagger}_q\}=-i \tanh{i2\theta},
\eea
whereas all other anti-commutators are equal to zero.
A set of operators $A$ and $B$ with canonical commutation relations and commuting among themselves is given  by
\bea
& A(\theta)\equiv \frac{\sqrt{[2]_q}}{2\sqrt 2} \big[ A_{q(\theta)} + A_{q(-\theta)} + A^{\dagger}_{q(\theta)} - A^{\dagger}_{q(-\theta)}\big], \\
& B(\theta)\equiv \frac{\sqrt{[2]_q}}{2\sqrt 2} \big[ B_{q(\theta)} + B_{q(-\theta)}  - B^{\dagger}_{q(\theta)} + B^{\dagger}_{q(-\theta)}\big].
\eea
and h.c., so that

\bea
\{A(\theta),A^{\dagger}(\theta)\}=1, \quad
\{B(\theta),B^{\dagger}(\theta)\}=1 , \quad
\{A(\theta),B^{\dagger}(\theta)\}=0.
\eea
and all other anti-commutators  equal to zero.
It's also possible to write:
\bea
A(\theta)=\frac{1}{\sqrt 2} \big(a(\theta) + \tilde{a}(\theta) \big), \qquad
 B(\theta)=\frac{1}{\sqrt 2} \big(a(\theta) - \tilde{a}(\theta) \big) .
\eea
with
\bea
\label{eqn:bogoliubov}
& a(\theta)=\frac{1}{\sqrt 2} \big(A(\theta) +B(\theta) \big)= a \cosh{i\theta} - \tilde{a}^{\dagger} \sinh{i \theta}, \\
\label{eqn:bogoliubov2}
& \tilde{a}(\theta)=\frac{1}{\sqrt 2} \big(A(\theta) -B(\theta) \big)= \tilde{a} \cosh{i\theta} - i a^{\dagger} \sinh{i \theta},
\eea
and
\bea
\label{eqn:anticommb}
&\{a(\theta),a^{\dagger}(\theta)\}=1, \quad
\{\tilde{a}(\theta),\tilde{a}^{\dagger}(\theta)\}=1 .
\eea
All others anti-commutators are equal to zero and $a(\theta)$ and $\tilde{a}(\theta)$ anti-commute among themselves.
Eqs.~(\ref{eqn:bogoliubov}) and (\ref{eqn:bogoliubov2}) are nothing but the Bogoliubov transformations of the pair of creation and annihilation operators $(a,\tilde{a})$ into a new pair $(a(\theta),\tilde{a}(\theta))$. In other words, Eqs.~(\ref{eqn:bogoliubov})-(\ref{eqn:anticommb}) show that the Bogoliubov-transformed operators $a(\theta)$ and $\tilde{a}(\theta)$ are linear combinations of the coproduct operators defined in terms of the deformation parameter $q(\theta)$ and their $\theta$-derivatives.

\section {Neutrino mixing}

In the context on NCSG, neutrinos appear as Majorana spinors. Therefore, we refer below to Majorana neutrinos; however, provided that convenient changes are introduced, our formalism can be readily extended to Dirac neutrinos and to other particle mixing.

Majorana fermions are self-conjugate particles. The charge-conjugation operator $C$  satisfies  the relations
\begin{equation}
C^{-1}\gamma_\mu C = -\gamma^T_\mu , \qquad C^{\dagger}=C^{-1} , \qquad C^T=-C .
\end{equation}
The charge conjugate $\psi^c$ of $\psi$ is defined by
\begin{equation}
\psi^c(x)\equiv\gamma_0 C \psi^\ast (x).
\end{equation}
A Majorana fermion is a field that satisfies both, the Dirac equation and the self-conjugation relation:
\begin{subequations}
\begin{align}
(i \slashed {\partial} - &m)\psi=0, \label{eqn:dirac}\\
&\psi=\psi^c . \label{eqn:majorana}
\end{align}
\end{subequations}
In NCSG, the neutrino mass terms in the Lagrangian is written as
\be \frac{1}{2}\sum_{\lambda\kappa}\bar\psi_{\lambda L}{\cal
S}_{\lambda\kappa}\hat\psi_{\kappa R}
+\frac{1}{2}\sum_{\lambda\kappa} \overline{\bar\psi_{\lambda L}{\cal
S}_{\lambda\kappa}\hat\psi_{\kappa R}}~. \ee
The subscripts $_{L,R}$ denote left-handed and right-handed
states, respectively. The off-diagonal elements of the symmetric matrix
${\cal S}_{\lambda\kappa}$ are the Dirac mass terms,  the
diagonal ones are the Majorana mass terms.

From the equations of motion one has that the largest
eigenvalue of the Majorana
mass matrix $M_R$ is of the order of the unification scale, while the
Dirac mass $M_\nu$ is of the order of the Fermi energy and thus much
smaller. Neutrino mixing and the seesaw mechanism are thus built in in the NCSG construction. We show below that neutrino mixing is implied by the doubling of the algebra, which is the core of Connes construction.

In order to proceed in our discussion, we first summarize briefly the QFT formalism for the neutrino mixing~\cite{threeNeutrinos}-\cite{BJV:2011}
%,neutrinoMixQFT,Blasone:2002jv,Palmer,}.
We present a number of relations which we need in order to show the role played in such a phenomenon by the Bogoliubov transformations. We will closely follow ref. \cite{GSV}, where a more detailed presentation is reported.

We start by introducing the Lagrangian:
\begin{equation}
L(x)=\bar{\psi_f}(x)(i \slashed {\partial} - M)\psi_f (x)=\bar{\psi_m}(x)(i \slashed {\partial} - M_d)\psi_m (x),
\end{equation}
with $\psi^T_f=(\nu_e,\nu_\mu)$ being the flavor fields and $M=\begin{pmatrix} m_e & m_{e\mu} \\ m_{e\mu} & m_{\mu} \end{pmatrix} $.

The flavor fields are connected to the free fields $\psi^T_m=(\nu_1,\nu_2)$, with $M_d~=~diag(m_1 , m_2)$, by the  the Pontecorvo mixing transformation
\bea \label{eqn:pontecorvo1}
& \nu_e (x)= \nu_1(x) \cos{\theta} + \nu_2 (x) \sin{\theta},\\
& \nu_{\mu} (x)= - \nu_1(x) \sin{\theta} + \nu_2 (x) \cos{\theta}.
\label{eqn:pontecorvo}
\eea
The free fields are given by
\begin{equation}
\nu_i (x) =\sum_{r=1,2} \int \frac {d^3 \bf{k}}{(2\pi)^\frac 3 2 } e^{i \bf{k\cdot x}}\big[u^r_{\bf{k},i}(t)\alpha^r_{\bf{k},i}+v^r_{-\bf{k},i}(t)\alpha^{r\dagger}_{-\bf{k},i}\big]\quad, \quad i=1,2
\end{equation}
where  $u^r_{\bf{k},i}(t)=e^{-i \omega_{\bf{k},i}t} u^r_{\bf{k},i}$ , $v^r_{\bf{k},i}(t)=e^{i \omega_{\bf{k},i}t} v^r_{\bf{k},i}$ and $\omega_{\bf{k},i}=\sqrt{\bf{k}^2+m^2_i}$ . In order for the Majorana condition \eqref{eqn:majorana} to be satisfied, the four spinors must also satisfy the following condition:
\begin{equation}
u^s_{\bf{k},i}=\gamma_0 C (v^s_{\bf{k},i})^\ast , \qquad v^s_{\bf{k},i}=\gamma_0 \, C (u^s_{\bf{k},i})^\ast .
\end{equation}
The equal-time anti-commutation relations are
\bea
&\{\nu^{\alpha}_i (x),\nu^{\beta\dagger}_j(y)\}_{t=t'}=\delta^3 (\bf{x-y}) \delta_{\alpha \beta} \delta_{i j} ,\\
& \{\nu^{\alpha}_i (x),\nu^{\beta}_j(y)\}_{t=t'}=\delta^3 (\bf{x-y}) (\gamma_0    C)_{\alpha \beta} \delta_{i j} ,
\eea
with $\alpha, \beta =1,\dots,4 $. And
\begin{equation}
\{\alpha^r_{-\bf{k},i},\alpha^{s\dagger}_{-\bf{q},i}\}_{t=t'}=\delta^3 (\bf{k_q}) \delta_{\bt{r s}} \delta_{i j} ,\qquad i=1,2 .
\end{equation}
All the other anti-commutators are zero. The ortonormality and completeness relations are :
\bea
& u^{r \dagger}_{{\bf k},i}  u^s_{{\bf k},i} = v^{r \dagger}_{{\bf k},i}  v^s_{{\bf k},i} = \delta{r s} , \qquad
u^{r \dagger}_{{\bf k},i}  v^s_{-{\bf k},i} = v^{r \dagger}_{-{\bf k},i}  u^s_{{\bf k},i} = 0 , \\
& \sum{r=1,2}  u^r_{{\bf k},i} u^{r \dagger}_{{\bf k},i}  + v^r_{{\bf k},i} v^{r \dagger}_{{\bf k},i}= {\bf 1}.
\eea
Eqs.~(\ref{eqn:pontecorvo1}) and (\ref{eqn:pontecorvo}) can be recast in the form :
\bea
& \nu^{\alpha}_e (x)= G^{-1}_{\theta}(t)\nu^{\alpha}_1(x) G_{\alpha}(t),\\
& \nu^{\alpha}_{\mu} (x)= G^{-1}_{\theta}(t)\nu^{\alpha}_2(x) G_{\alpha}(t),
\eea
where $G_{\theta}(t)$ is given by
\begin{equation}
G_{\theta}(t)=\exp{\Big[\frac \theta 2 \int  d^3\bf{x} \big(\nu^{\dagger}_1(x) \nu_2(x) - \nu^{\dagger}_2(x) \nu_1(x) \big)}\Big] .
\end{equation}
One has $G_{\theta}(t)=\textstyle\prod_{\bf{k}} G^{\bf{k}}_{\theta}(t)$ . For a given $\bf{k}$ , in the reference frame where $\bf{k} =(0 , 0 , |\bf{k}|)$ , the spins decuple and one has $G^{\bf{k}}_{\theta}(t)=\textstyle\prod_r  G^{\bf{k}, r}_{\theta}(t)$ with
\begin{equation}
G^{\bf{k}}_{\theta}(t)=\exp{\Big\{ \theta \Big[ U^\ast_{\bf{k}}(t) \alpha^{r\dagger}_{\bf{k},1} \alpha^{r}_{\bf{k},2} - U_{\bf{k}}(t) \alpha^{r\dagger}_{-\bf{k},2} \alpha^{r}_{-\bf{k},1}-\epsilon^r V^\ast_{\bf{k}}(t) \alpha^{r}_{-\bf{k},1} \alpha^{r}_{\bf{k},2} +\epsilon^r V_{\bf{k}}(t) \alpha^{r\dagger}_{\bf{k},1} \alpha^{r\dagger}_{-\bf{k},2} \Big]\Big\}} ,
\end{equation}
where $U_{\bf{k}}(t)$ e $V_{\bf{k}}(t)$ are Bogoliubov coefficients given by
\begin{equation}
 U_{\bf{k}}(t)\equiv |U_{\bf{k}}|e^{i(\omega_{\bf{k},2}-\omega_{\bf{k},1})t}\qquad,\qquad V_{\bf{k}}(t)\equiv |{V_{\bf{k}}}| e^{i(\omega_{\bf{k},2}+\omega_{\bf{k},1})t}\qquad,
\end{equation}
\begin{equation}
|{U_{\bf{k}}}|\equiv \Big(\frac {\omega_{\bf{k},1}+m_1}{2\omega_{\bf{k},1}} \Big)^\frac 1 2  \Big(\frac {\omega_{\bf{k},2}+m_2}{2\omega_{\bf{k},2}} \Big)^\frac 1 2  \Big(1 +  \frac {|{\bf{k}}|^2}{(\omega_{\bf{k},1}+m_1)(\omega_{\bf{k},2}+m_2)} \Big) ,
\end{equation}
\begin{equation}
|{V_{\bf{k}}}|\equiv \Big(\frac {\omega_{\bf{k},1}+m_1}{2\omega_{\bf{k},1}} \Big)^\frac 1 2  \Big(\frac {\omega_{\bf{k},2}+m_2}{2\omega_{\bf{k},2}} \Big)^\frac 1 2  \Big(\frac {|{\bf{k}}|}{(\omega_{\bf{k},2}+m_2)} - \frac {|{\bf{k}}|}{(\omega_{\bf{k},1}+m_1)} \Big) ,
\end{equation}
\begin{equation}
|{U_{\bf{k}}}|^2 + |{V_{\bf{k}}}|^2 =1.
\end{equation}
The flavor fields can be thus expanded as :
\begin{equation}
\nu_\sigma (x) =\sum_{r=1,2} \int \frac {d^3 \bf{k}}{(2\pi)^\frac 3 2 } e^{i \bf{k\cdot x}}\big[u^r_{\bf{k},j}(t)\alpha^r_{\bf{k},\sigma}+v^r_{-\bf{k},j}(t)\alpha^{r\dagger}_{-\bf{k},\sigma}\big]\quad,
\end{equation}
with $\sigma , j = (e, 1),(\mu , 2)$ and the flavor annihilation operators given by (for $\bf{k} =(0 , 0 , |\bf{k}|)$):
\bea
\label{eqn:opcrea1}
& \alpha^r_{\bf{k},e}\equiv G^{-1}_{\theta}(t)\,\alpha^r_{-\bf{k},1} G_{\theta}(t)= \cos{\theta}\,\alpha^r_{\bf{k},1} + \sin{\theta} \big(U^\ast_{\bf{k}}(t)\,\alpha^r_{\bf{k},2} +\epsilon^r V^\ast_{\bf{k}}(t)\,\alpha^{r\dagger}_{-\bf{k},2} \big),\\ \label{eqn:opcrea2}
& \alpha^r_{\bf{k},\mu}\equiv G^{-1}_{\theta}(t)\,\alpha^r_{-\bf{k},2} G_{\theta}(t)=  \cos{\theta}\,\alpha^r_{\bf{k},2} - \sin{\theta} \big(U_{\bf{k}}(t)\,\alpha^r_{\bf{k},1} +\epsilon^r V_{\bf{k}}(t)\,\alpha^{r\dagger}_{-\bf{k},1} \big).
\eea
Notice the presence of the Bogoliubov coefficients $U_{\bf{k}}(t)$ and $V_{\bf{k}}(t)$ in Eqs.~(\ref{eqn:opcrea1}) and (\ref{eqn:opcrea2}). {\it These relations clearly show how Bogoliubov transformations enter the neutrino mixing transformations}.

Next, we consider the action of the mixing generator  on the vacuum $|0\rangle_{1,2}$. The flavor vacuum is defined as:
\begin{equation}
|0(\theta,t)\rangle_{e,\mu}\equiv G^{-1}_{\theta}(t)|0\rangle_{1,2}
\end{equation}
The state for a mixed particle with definite flavor, spin and momentum is given by:
\begin{equation}
|{ \alpha^r_{\bf{k},e}}\rangle \equiv \alpha^{r\dagger}_{\bf{k},e} |{0(\theta,t)}\rangle_{e,\mu}= G^{-1}_{\theta}(t) \alpha^{r\dagger}_{\bf{k},1}|{0}\rangle_{1,2}\quad .
\end{equation}

The anti-commutators of the flavor ladder operators at different times are then computed and
the following quantity is found to be constant in time:
\begin{equation}
|{ \{ \alpha^r_{\bf{k},e}(t), \alpha^{r\dagger}_{\bf{k},e}(t')\}}|^2 + {|{ \{ \alpha^{r\dagger}_{-\bf{k},e}(t), \alpha^{r\dagger}_{\bf{k},e}(t')\} }}|^2 + {|{ \{ \alpha^r_{\bf{k},\mu}(t), \alpha^{r\dagger}_{\bf{k},e}(t')\} }}|^2 + {|{ \{ \alpha^{r\dagger}_{-\bf{k},\mu}(t), \alpha^{r\dagger}_{\bf{k},e}(t')\}}}|^2 = 1
\end{equation}
The energy-momentum tensor for the fermion field is defined by ${\cal J}^{\mu \nu} \equiv i \bar{\psi}\gamma^{\nu}\partial_{\mu}\psi$  and  the momentum operator is given by $P^j \equiv \int d^3 \bf{x} \quad {\cal J}^{0j}(x)$. For the free fields we have:
\begin{equation}
\bf{P}_i= \int d^3 \bf{x} \psi^\dagger_i (x) (- i \nabla) \psi_i (x) = \int d^3 \bf{k} \sum_{r=1,2} \bf{k} \Big(\alpha^{r\dagger}_{\bf{k},i}  \alpha^{r}_{\bf{k},i} -  \alpha^{r\dagger}_{-\bf{k},i}  \alpha^{r}_{-\bf{k},i} \Big),
\end{equation}
for $i = 1,2$. The momentum operator for mixed fields is given by:
\begin{equation}
\bf{P}_\sigma (t)= \int d^3 \bf{x} \psi^\dagger_\sigma (x) (- i \nabla) \psi_\sigma (x) = \int d^3 \bf{k} \sum_{r=1,2} \bf{k} \Big(\alpha^{r\dagger}_{\bf{k},\sigma}(t) \alpha^{r}_{\bf{k},\sigma}(t) -  \alpha^{r\dagger}_{-\bf{k},\sigma}(t)  \alpha^{r}_{-\bf{k},\sigma}(t) \Big),
\end{equation}
for $\sigma = e,\mu$. We have $\bf{P}_\sigma (t)=G^{-1}_{\theta}(t) \bf{P}_i G_{\theta}(t) $ and the conservation of total momentum:
\begin{equation}
\bf{P}_e(t)+\bf{P}_\mu(t)= \bf{P}_1 + \bf{P}_2 \equiv \bf{P} \quad, \quad [\bf{P}, G_{\theta}(t)]=0\quad,\quad [\bf{P}, H]=0 .
\end{equation}
At time $t=0$, the flavor state $|{ \alpha^r_{\bf{k},e}}\rangle \equiv |{ \alpha^r_{\bf{k},e}(0)}\rangle $ is an eigenstate of the momentum operator $\bf{P}_e(0)$:
\begin{equation}
\bf{P}_e(0)|{ \alpha^r_{\bf{k},e}}\rangle \equiv \bf{k} |{ \alpha^r_{\bf{k},e}}\rangle .
\end{equation}
At time $t\ne0$ the expectation value of the momentum in such a state (normalized to initial time value) gives~\cite{Palmer}:
\begin{equation}
{\cal P}^e_{\bf{k},\sigma}(t)\equiv \frac {\langle{ \alpha^r_{\bf{k},e}}| \bf{P}_\sigma(t)|{ \alpha^r_{\bf{k},e}}\rangle }{\langle{ \alpha^r_{\bf{k},e}}| \bf{P}_\sigma(0)|{ \alpha^r_{\bf{k},e}} \rangle}= |{ \{ \alpha^r_{\bf{k},e}(t), \alpha^{r\dagger}_{\bf{k},e}(t')\}}|^2 + {|{ \{ \alpha^{r\dagger}_{-\bf{k},e}(t), \alpha^{r\dagger}_{\bf{k},e}(t')\} }}|^2 ,
\end{equation}
for $\sigma=e,\mu$, and the flavor vacuum expectation value of the momentum operator $\bf{P}_\sigma(t)$ vanishes at all time :
\begin{equation}
{}_{e,\mu}\langle{0} |\bf{P_\sigma(t)}|{0}_{e,\mu}\rangle \quad, \qquad \sigma=e,\mu .
\end{equation}
The explicit calculation of the oscillating quantities ${\cal P}^e_{\bf{k},\sigma}(t)$ is found to be~\cite{Palmer}:
\begin{equation}
{\cal P}^e_{\bf{k},e}(t)= 1-\sin^2{2\theta} \Big[ |{U_{\bf{k}}}| \sin^2{\frac {\omega_{k,2} -\omega_{k,1} }{2} t} +|{V_{\bf{k}}}| \sin^2{\frac {\omega_{k,2} +\omega_{k,1} }{2} t} \Big],
\end{equation}
\begin{equation}
{\cal P}^e_{\bf{k},\mu}(t)= \sin^2{2\theta} \Big[ |{U_{\bf{k}}}| \sin^2{\frac {\omega_{k,2} -\omega_{k,1} }{2} t} +|{V_{\bf{k}}}| \sin^2{\frac {\omega_{k,2} +\omega_{k,1} }{2} t} \Big].
\end{equation}
These formulas give the neutrino oscillation formulas in the QFT formalism and are different from the quantum mechanical approximation provided by the Pontecorvo formulas, to which they reduce in the large $|\bf k|$ limit, namely in $|{U_{\bf{k}}}| \rar 1$ limit. In such a limit the Bogoliubov transformations and the algebra doubling are washed away. But then, as well, there is no NCSG construction anymore.

\section{Conclusion}

The study of the neutrino mixing carried on in the QFT formalism has
shown that the mixing transformation is not just a simple rotation,
but a transformation made of a rotation ``nested'' with a Bogoliubov
transformation. This makes the mass vacuum and the flavor vacuum
unitarily inequivalent. The Bogoliubov transformation is in turn the
result of the algebra doubling ${\cal A} = {\cal A}_1 \otimes {\cal
  A}_2 $ acting on the space ${\cal H} = {\cal H}_1 \otimes {\cal
  H}_2$, as we have discussed in Section 2. Then we have shown that
Bogoliubov transformations are built in in the algebra doubling in the
NCSG construction. In fact the doubling of the algebra is a
characterizing feature at the basis of the NCSG construction. The
neutrino mixing mechanism appears thus naturally incorporated in the
NCSG construction. Since Majorana neutrinos appear in the NCSG
construction, we consider their mixing and show how the neutrino
oscillations emerge. We also show that in the limit of large $|\bf k|$
the quantum mechanical approximation of the Pontocorvo mixing is
obtained and the Bogoliubov transformation reduces to the identity
transformation. In such a limit the doubling of the algebra,
characterizing feature of the NCSG construction, is also lost.

%%%%%%%%%%%%%%%%%%%%%%%%%%%%%%%%%%%%%%%%%%%
\section*{References}
%\bibliography{iopart-num}

\end{document}